\documentclass{article}
\usepackage{spconf,amsmath,graphicx,hyperref,dirtytalk}
\usepackage{adjustbox}
\usepackage{booktabs}
\usepackage{float}
\usepackage{textcomp}
\usepackage{algorithm}
\usepackage{algpseudocode}
\usepackage{subfig}
\usepackage{bm}
\usepackage{amssymb}
\usepackage{color}
\usepackage{tikz}
\usepackage{anyfontsize}
\usetikzlibrary{shapes}
\newcommand{\ra}[1]{\renewcommand{\arraystretch}{#1}}

\hypersetup{hidelinks}

\newcommand{\markertriang}{\raisebox{0.5pt}{\tikz{\node[draw,scale=0.3,regular polygon, regular polygon sides=3,fill=white!10!cyan,rotate=0](){};}}}
\title{NOMAD: Unsupervised Learning of Perceptual Embeddings for Speech Enhancement and Non-matching Reference Audio Quality Assessment}
\name{Alessandro Ragano$^{1}$, \qquad Jan Skoglund$^{2}$, \qquad Andrew Hines$^{1}$ \thanks{This work has emanated from research supported in part by the Google and research grants from Science Foundation Ireland (SFI) co-funded under the European Regional Development Fund under Grant Number
13/RC/2289\_P2.}}

\address{$^{1 }$School of Computer Science, University College Dublin, Ireland,  $^{2}$Google LLC, San Francisco, USA}

\begin{document}
\ninept
\maketitle
\begin{abstract}
This paper presents NOMAD (\underline{No}n-\underline{M}atching \underline{A}udio \underline{D}istance), a differentiable perceptual similarity metric that measures the distance of a degraded signal against non-matching references. The proposed method is based on learning deep feature embeddings via a triplet loss guided by the Neurogram Similarity Index Measure (NSIM) to capture degradation intensity. During inference, the similarity score between any two audio samples is computed through Euclidean distance of their embeddings. NOMAD is fully unsupervised and can be used in general perceptual audio tasks for audio analysis e.g. quality assessment and generative tasks such as speech enhancement and speech synthesis. 
The proposed method is evaluated with 3 tasks. Ranking degradation intensity, predicting speech quality, and as a loss function for speech enhancement. Results indicate NOMAD outperforms other non-matching reference approaches in both ranking degradation intensity and quality assessment, exhibiting competitive performance with full-reference audio metrics. NOMAD demonstrates a promising technique that mimics human capabilities in assessing audio quality with non-matching references to learn perceptual embeddings without the need for human-generated labels.
\end{abstract}
\vspace{-2mm}
\section{Introduction}
\label{sec:intro}

Objective speech and audio quality assessment techniques include \textit{full-reference} metrics~\cite{hines2015visqol,rix2001perceptual,manocha2021cdpam,jassim2021warp}, using both degraded and clean signals, and \textit{no-reference} metrics~\cite{avila2019non,ragano2021more,Quality-Net,serra2021sesqa,catellier2020wawenets} that predict mean opinion scores (MOS) from the degraded signal only. No-reference metrics overcome issues of full-reference metrics, like sensitivity to imperceptible differences between degraded and reference signals~\cite{jassim2021warp}, as well as the lack of need for a reference signal. However, \mbox{no-reference} metrics assume absolute quality, as MOS is given without a reference, using the absolute category rating (ACR) scale~\cite{ITUP800}, which is calibrated with anchors. Yet, MOS distributions remain relative due to biases~\cite{zielinski2008some} and stimulus dependence. We observe that merging MOS databases for no-reference metrics is uncommon due to label space differences; MOS of $4.0$ has different meanings across databases. 
In ~\cite{manocha2021noresqa} it has been noticed that no-reference models would need to learn the hidden references used by raters when judging quality which can be very challenging. To solve this,~\cite{manocha2021noresqa} proposed NORESQA which measures the perceived quality of a degraded signal against ~\textit{non-matching references} i.e. using any clean speech signal, not necessarily the clean counterpart of the degraded signal. The advantage of non-matching references is twofold: the clean counterpart is not required and quality can be measured relatively to any other signal. If any clean speech is used as a non-matching reference, then absolute quality is measured. This approach reflects the higher capacity of humans in sensory judgement when comparing stimuli instead of absolute quality~\cite{lawless2010sensory}.

In this work, we introduce NOMAD (\underline{No}n-\underline{M}atching \underline{A}udio \underline{D}istance), a perceptual differentiable audio metric that operates with any non-matching reference. Our method creates an embedding space where signals with similar degradation intensity are close. We employ the triplet loss~\cite{schroff2015facenet}, a popular contrastive approach in computer vision for metric learning, to achieve this. We use degradation intensity as a label, which is linked to quality, and programmatically controlled without relying on human labels. However, the challenge with degradation intensity parameters is their lack of comparability across different degradations. To address this, we propose to use the Neurogram Similarity Index Measure (NSIM)~\cite{hines2012speech}, a spectro-temporal similarity between degraded and clean signals ranging from $0.0$ to $1.0$. A non-matching reference metric must be reference-invariant, consistently producing the same score for a degraded signal regardless of the clean signal used for comparison. We attain this by training a feature space invariant to speaker and sentence characteristics, using the self-supervised learning (SSL) model wav2vec 2.0~\cite{baevski2020wav2vec}, which has proven efficacy across diverse downstream tasks with distinct variational factors~\cite{becerra22_interspeech,makiuchi2021multimodal,ravi22_interspeech}.

NOMAD can be used in diverse applications: quality prediction, perceptual audio retrieval, parallel and non-parallel speech enhancement, and waveform synthesis like text-to-speech. We evaluate NOMAD's performance in three tasks: ranking degradation intensity, speech quality assessment, and speech enhancement training loss. The PyTorch code, pip package, and dataset generation code for training and validation are available on GitHub\footnote{\url{https://github.com/alessandroragano/nomad}}.

\vspace{-4mm}
\section{Proposed Method}
Our approach relies on the assumption that audio quality is linked to degradation intensity. We aim to develop a similarity metric that captures degradation intensity in degraded audio, irrespective of factors like speaker or sentence attributes in speech. Let us start by considering a scenario with a single degradation type, such as background noise. We can model degradation using a function denoted as $h(\cdot, \alpha)$. When applied to clean speech $\bm{x}$, this function generates a signal with degradation intensity depending on a scalar parameter, $\alpha$ e.g., SNR.
Given two values $\alpha_i$ and $\alpha_j$, along with the corresponding degraded samples $\bm{x}_i = h(\bm{x}, \alpha_i)$ and $\bm{x}_j = h(\bm{x}, \alpha_j)$, the degradation parameter can be used as a label\footnote{The direction of $\alpha$ depends on the type of degradation.} to learn a similarity function $\mathcal{D}(\cdot,\cdot)$ that follows the constraint: 
\begin{equation} \label{eq:single_degr}
 \alpha_i > \alpha_j \implies \mathcal{D}(\bm{x},\bm{x}_i) > \mathcal{D}(\bm{x}, \bm{x}_j)
\end{equation}
The idea is to induce a semantic order in the feature space based on the level of degradation. Using only one degradation is limiting for generalization. A perceptual audio similarity metric should be able to capture information from multiple degradations. Let us consider the case where the clean speech $\bm{x}$ is perturbed with two different degradations producing two signals $\bm{x}_a = h_a(\bm{x}, \alpha)$ and $\bm{x}_b = h_b(\bm{x}, \beta)$. In this scenario, each degradation is controlled by a single scalar parameter and it is not possible to establish an order between the two parameters such that $\alpha > \beta \implies \mathcal{D}(\bm{x},\bm{x}_a) > \mathcal{D}(\bm{x}, \bm{x}_b). $

In order to establish cross-degradation similarity with respect to to the clean reference we propose to use the NSIM~\cite{hines2012speech} which is a spectro-temporal measure of similarity between a degraded signal and its clean counterpart and it has been proven to model human speech quality perception~\cite{hines2015visqol}. The NSIM is a score between $0.0$ and $1.0$ relative to the reference signal defined as:
\begin{equation} \label{eq:nsim}
    Q(r,d) = \frac{2\mu_r\mu_d + C_1}{\mu_{r}^2 + \mu_{d}^2 + C_2} \cdot \frac{\sigma_{rd} + C_3}{\sigma_r\cdot\sigma_d + C_3}.
\end{equation}
Here, $r$ represents the clean speech spectrogram, and $d$ denotes the degraded spectrogram. The NSIM relies on statistical measures: $\mu_r$ and $\sigma_r$ are the mean and standard deviation of the reference spectrogram, while $\mu_d$ and $\sigma_d$ are the mean and standard deviation of the degraded spectrogram. Additionally, $\sigma_{rd}$ denotes the cross-correlation between the reference and degraded spectrogram. The constant values $C_1 = 0.01L$ and $C_2=C_3=(0.03L)^2$ are determined based on the intensity range $L$ of the reference spectrogram and used for boundary conditions~\cite{hines2015visqol}. 

By employing the NSIM, we gain the capability to compare multiple degradations, leading to the formulation of Equation~\ref{eq:nsim_distance}:
\begin{equation} \label{eq:nsim_distance}
NSIM_a > NSIM_b \implies \mathcal{D}(\bm{x},\bm{x}_a) < \mathcal{D}(\bm{x}, \bm{x}_b)
\end{equation}
In this equation, $\bm{x}_a$ denotes the signal obtained using degradation $a$, while $\bm{x}_b$ is obtained from another degradation $b$. The constraint implies that $\bm{x}_a$ must be closer to $\bm{x}$ than $\bm{x}_b$ since the NSIM of degradation $a$ is higher.
In the following section, we illustrate how the NSIM can be leveraged to learn a perceptual distance function that is cross-reference, even though the NSIM is a score relative to the same reference signal.

\vspace{-3mm}
\subsection{Loss Function}
To model a perceptual similarity metric as formulated in Equation \ref{eq:nsim_distance} we employ the triplet margin loss function~\cite{schroff2015facenet}, supervised by the order of the NSIM scores. The triplet loss function is commonly used for deep metric learning and it is formulated as follows:
\small
\begin{equation} \label{eq:triplet_loss}
 \sum_{i}^{N} \max \left(0, ||f(\bm{x}_i^a) - f(\bm{x}_i^p) ||_2^2 - ||f(\bm{x}_i^a) - f(\bm{x}_i^n) ||_2^2 + m\right)
\end{equation}
Here, $\bm{x}_i^a$, $\bm{x}_i^p$, and $\bm{x}_i^n$ represent anchor, positive, and negative samples respectively and $f(\cdot)$ is the neural network producing the embeddings.
The goal of the triplet loss is to make the squared Euclidean distance between the anchor-positive pair smaller and increase the distance between the anchor-negative pair by a specified margin $m$.

The training process includes triplet sampling. As we lack categorical labels like $y^a=y^p$ and $y^a\neq y^n$ for relationships, we observe that the concept of \say{closeness} is relative in regression problems. For instance, someone $180$ cm tall is closer to $170$ cm than $200$ cm. Similarly, an NSIM of $0.90$ is closer to $0.88$ than to $0.99$.
Our method employs NSIM space to represent "closeness" between speech samples. This develops feature embeddings capturing similarities among speech samples with similar NSIM and thus close degradation intensity.  At test time, we measure the similarity score between embeddings using the Euclidean distance. The challenge of triplet sampling, crucial for the triplet loss to work~\cite{schroff2015facenet}, is addressed in the next section.
Our approach does not aim to predict exact NSIM scores as they are relative to the reference signal used. Predicting NSIM would prevent comparing signals from different sources.

\begin{figure}[!t]
\centering
\includegraphics[width=0.80\linewidth]{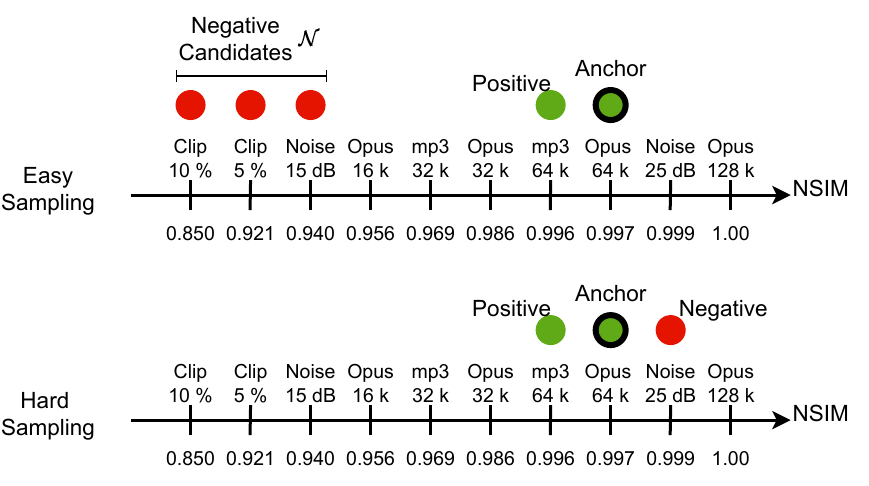}
\caption{Easy sampling strategy (above). The conditions that have distance $|Q_{k,m} - Q^a|$ lower than $|Q^p- Q^a| + s$ are excluded. Hard sampling strategy (below). The negative is the one with the shortest distance from the anchor after the positive. }
\label{fig:sampling_strategy}
\vspace{-5mm}
\end{figure}

\vspace{-3mm}
\subsection{Sampling Strategy}
Large batch sizes are required to find harder triplets in the embedding space e.g. 1800~\cite{schroff2015facenet}.
To avoid memory issues with large deep models we combine two different sampling strategies called easy and hard sampling not requiring large batch sizes since we do not use the embedding space. Initial experiments yielded better results than using only one of them. Our strategy is based on the idea that harder triplets can be identified with the NSIM. Intuitively, the larger the NSIM between two samples, the easier is the task since the same reference is used in the triplet.

For a clean speech file $\bm{x}$, we consider the sample set $\mathcal{P}_x = {(\bm{x}_{k,m}, Q_{k,m})}$ which includes degraded versions of $\bm{x}$ perturbed by $M$ degradations at $K$ levels and their corresponding NSIM value $Q_{k,m}$. We sample a clean file $\bm{x}$ and an anchor $(\bm{x}^a, Q^a)$ from $\mathcal{P}_x$. The positive is the sample with the closest NSIM score to the anchor: $Q^p=\operatorname*{arg,min}{(k,m)} |Q_{k,m} - Q^a|$. Easy and hard sampling differ in negative selection. In easy strategy, the negative is sampled from $\mathcal{N} = \{(k,m), |Q_{k,m} - Q^a| > |Q^a - Q^p| + s \}$. The set of negative samples $\mathcal{N}$ includes all the samples of $\mathcal{P}_x$ where the NSIM scores are more distant from the anchor $Q^a$ than the positive $Q^p$ by at least a margin $s$.
Hard approach picks the closest sample to anchor after the positive: $Q^n=\operatorname*{arg,min}{(k,m)} |Q_{k,m} - Q^a| > |Q^p - Q^a|$ which is the hardest negative to contrast. See Figure \ref{fig:sampling_strategy} for easy and hard sampling illustrations.

\vspace{-3mm}
\subsection{Architecture}
\begin{figure}[!t]
\centering
\includegraphics[width=0.74\linewidth]{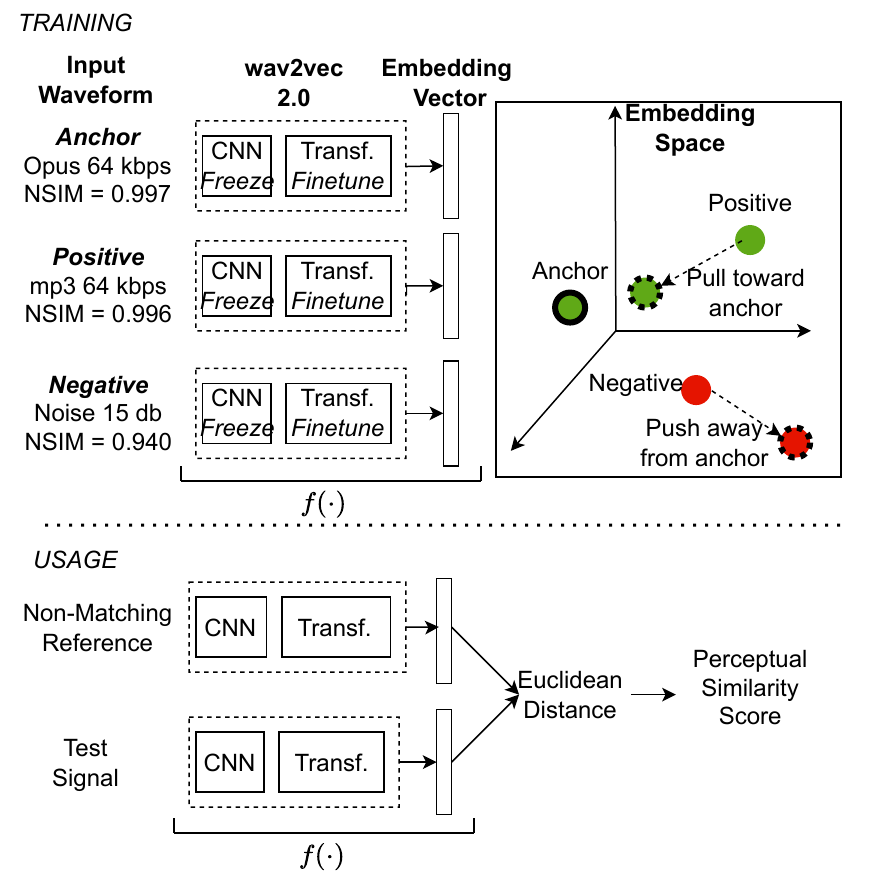}
\caption{Overview of the proposed method NOMAD.}
\label{fig:nomad_overview}
\vspace{-5mm}
\end{figure}

In the method outlined, each triplet uses a distinct reference, but within each triplet, the anchor, positive, and negative samples all come from the same reference (Figure \ref{fig:nomad_overview}). Contrasting degraded samples from the same reference rather than dissimilar ones during training helps create an embedding space that captures degradation levels, facilitating the use of non-matching references. 

To illustrate this, consider a triplet with the same degradation, like adding background noise linearly to clean speech $\bm{x}$ with intensity $\alpha$, yielding noisy speech $\bm{y} = \bm{x} + \alpha\bm{s}$.
Here, $\alpha$ has three distinct scalar parameters: $a$ for the anchor, $p$ for the positive, and $n$ for the negative example. The goal is to obtain an embedding space where content and degradation are disentangled as they are in the waveform space, making Equation~\ref{eq:triplet_loss}:
 $|| (\bm{x} + a\bm{s}) -  (\bm{x} + p\bm{s}) ||_2^2  - || (\bm{x} + a\bm{s}) - (\bm{x} + n\bm{s}) ||_2^2 + m.$
This objective is facilitated if we use the same clean speech $\bm{x}$. Indeed, during training the model is forced to cancel out the clean component which is the common part between the 3 signals and to rely on the residual between both pairs anchor-positive and anchor-negative respectively to minimize the loss.
To achieve this, a feature representation model is needed that can disentangle factors like content and degradation, ensuring that samples with similar degradation levels are close in the embedding space. Attenuating the clean component is not trivial since degradations are usually more complex than a sum between two signals e.g. convolution in reverberated speech.
To this end, we propose using the pre-trained \texttt{BASE} wav2vec 2.0 model~\cite{baevski2020wav2vec}. It consists of 7 convolutional layers followed by 12 transformer layers, yielding a $768$-dimensional feature vector per time frame. We take the average over the time dimension at the final transformer layer, followed by a ReLU + $256$-dimensional embedding layer. Embeddings are L2 normalized as described in~\cite{schroff2015facenet}.
We emphasize that constructing triplets from the same reference and utilizing the pre-trained wav2vec 2.0 model are crucial to achieving the results shown below. We tested models built from scratch and triplets with negative examples from different references, but both led to decreased performance.

\vspace{-4mm}
\subsection{Usage}
\sloppy NOMAD embeddings can be used as follows. Given a degraded recording $x_{test}$ and a non-matching clean reference $x_{nmr}$ we calculate the euclidean distance between the embeddings as 
$NOMAD(x_{test}, x_{nmr}) = || f(x_{test}) - f(x_{nmr}) ||_2 $ where $f(\cdot)$ is the model producing perceptual embeddings. NOMAD scores may vary based on the reference used. To minimize variability we calculate the mean on a large set of $I$ non-matching references $x_{nmr}^i$ as follows: $NOMAD(x_{test}, x_{}) = \frac{1}{I}   \sum_{i=1}^{I} NOMAD(x_{test}, x_{ref}^i)$. 

\section{Performance Evaluation}
\subsection{Experimental Setup}
Training and validation sets of NOMAD are created from the Librispeech~\cite{panayotov2015librispeech} partition \texttt{train-clean-100} which consists of $\approx 100$ hours of English clean speech spoken by $125$ female speakers and $126$ male speakers and recorded at $16$ kHz. We choose $M=4$ perturbations: speech clipping, background noise, Opus, and mp3 codecs. Each perturbation is generated at $K=5$ levels. Speech clipping is generated by choosing the percentage of samples to clip in the waveforms with 5\%, 10\%, 25\%, 40\%, 60\%. Background noise is controlled with the amount of noise injected in the clean signal with 0, 8, 15, 25, and 40 db SNR. Noise files are randomly extracted from the training set of the \mbox{MS-SNSD} dataset~\cite{reddy2019scalable}. Speech codecs mp3 and Opus are generated with the following conditions: 8, 16, 32, 64, and 128 kbps.
Both easy and hard sampling subsets are created using $\approx 8000$ triplets which are split into $80\%$ training and $20\%$ validation. Training and validation do not overlap in terms of clean speech sources and they include the same conditions.
For the easy sampling we set the hyperparameter $s=0.05$ to avoid negative samples that are too close to the positive as illustrated in Figure~\ref{fig:sampling_strategy}. The margin $m$ in the triplet loss is set to $0.2$.
The NSIM values are calculated from the ViSQOL v3 model~\cite{chinen2020visqol} which outputs patchwise scores. To get utterance-level scores, the average of all patch NSIM scores is computed.
During training we freeze the convolutional layers, finetune the transformer layers with a learning rate equal to 0.00001 and the embedding layer with a learning rate set to 0.0001. Both learning rates decay exponentially with a decay factor of $0.9$ every $20$ epochs without improvement. The batch size is set to 8.
Training is stopped when the triplet loss does not decrease on the validation set for 200 epochs.
\begin{figure}[!t]
\centering
\includegraphics[width=0.86\linewidth]{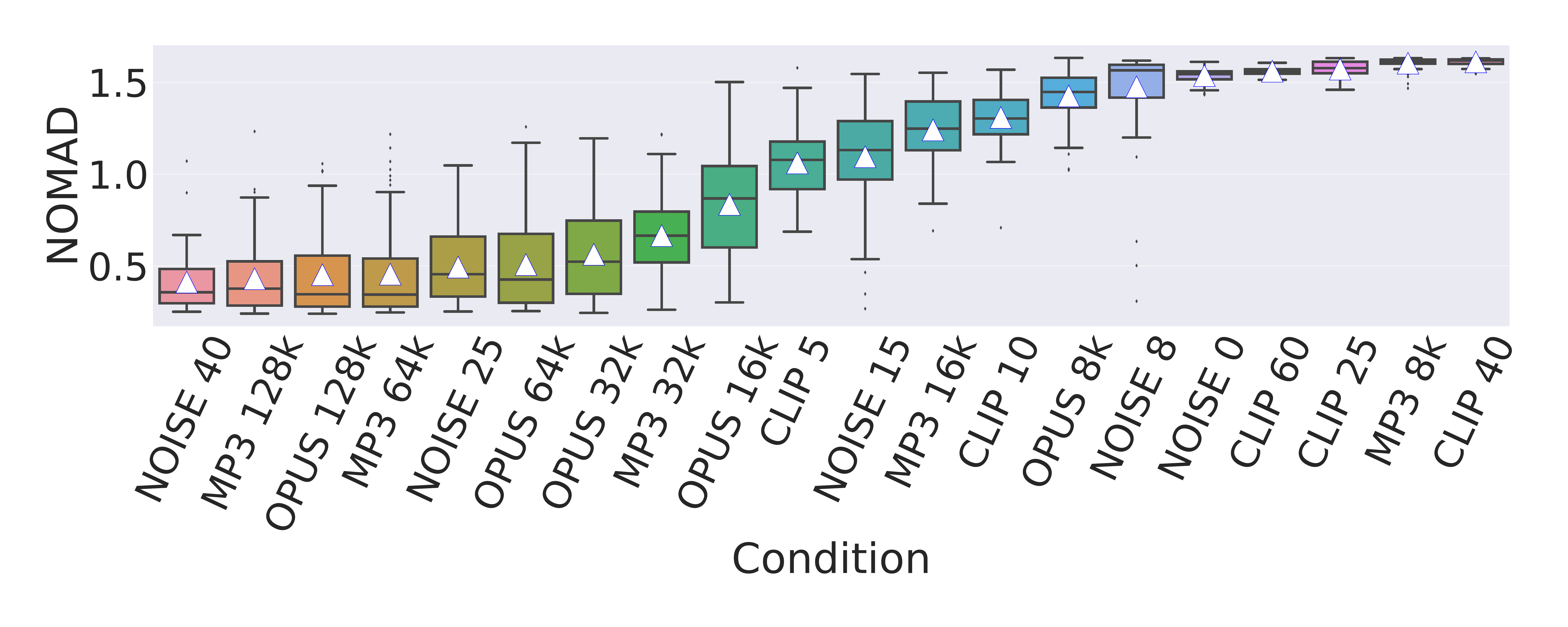}
\caption{Validation set conditions sorted by the NOMAD scores (\protect\markertriang).}
\label{fig:degradation_ranking}
\vspace{-3mm}
\end{figure}
All the non-matching reference model scores are calculated using a sample of $\approx 900$ clean speech sources from the TSP database~\cite{kabal2002tsp}. Recordings are downsampled to $16$ kHz and the $4$ speakers that are used in the TCD-VoIP database~\cite{harte2015tcd} are excluded since it is one of the test databases that we use below. 

\vspace{-6pt}
\subsection{Degradation Monotonicity}
A descriptive examination is done on the validation set, depicted in Figure~\ref{fig:degradation_ranking}. Here, we display averaged NOMAD scores, ordered from low (closer to non-matching clean speech) to high (more distorted). NOMAD ranks all validation conditions except clipping, demonstrating its performance and adaptability to non-matching references.
We investigate unseen conditions and degradations. We assess degradation monotonicity concerning intensity and quality using Spearman's rank correlation coefficient (SC). For ranking degradation parameters, we create an artificial test set from the Librispeech partition \texttt{test-clean}. This includes 26 unseen conditions for mp3 and Opus, 20 for clip, and 25 for background noise, drawn from the MS-SNSD~\cite{reddy2019scalable} test set. To assess out-of-domain degradations, we create 30 conditions of reverberated speech using the SoX audio effects library~\cite{sox} and 6 conditions using the Vorbis codec. Every degraded sample is created from a distinct clean source file. Scores are calculated using clean speech sources from the TSP database as non-matching references.
We compare NOMAD with 2 baselines; the average over the last transformer layer of the pre-trained \texttt{BASE} model wav2vec 2.0 and NORESQA, summarized in Table~\ref{tab:ranking}. Results show NOMAD outperforms in most conditions, except clipping, where wav2vec 2.0 ranking is better. This highlights wav2vec 2.0's suitability as a pre-trained model for NOMAD, with our approach contributing to NOMAD's superior performance.

While ranking by degradation intensity has its limitations, as it may not always reflect perception, we conduct a degradation-wise evaluation against MOS using the TCD-VoIP database, which includes both seen and unseen degradations. Table~\ref{tab:ranking} confirms NOMAD's perceptual ranking ability, surpassing wav2vec 2.0 and NORESQA in all degradations except background noise.

\begin{table}[!t]
\caption[C]{SC using degradation intensity and quality.}
\centering
\Huge
\ra{1.2}
\begin{adjustbox}{max width=0.49\textwidth}
\fontsize{94pt}{94pt}\selectfont
\begin{tabular}{@{}lccccccccccc|ccccccccccccccccccccc@{}}\toprule

& \multicolumn{11}{c} {\textbf{Ranking Intensity}} \vline & \multicolumn{10}{c}{\textbf{Ranking Quality, TCD-VoIP}}\\

&  NOISE && OPUS && MP3 && CLIP && VORB. && REV. && CLIP && NOISE && ECHO && CHOP && CSPKR  \\ \midrule
NOMAD   & \textbf{-0.74 }&& \textbf{-0.68} &&\textbf{ -0.73} && 0.89 && \textbf{-0.83} && \textbf{0.89} && \textbf{-0.98} && -0.70 && \textbf{-0.84} && \textbf{-0.86} && \textbf{-0.82}  \\
w2v     & -0.73 && -0.42 && -0.54 && \textbf{0.92} && 0.03 && 0.87 && -0.93 && \textbf{-0.79} && -0.76 && -0.33 && -0.66 \\
NORESQA & -0.41 && -0.20 && -0.45 && 0.64 && -0.77 && 0.81 && -0.52 && -0.18 && -0.01 && -0.37 && -0.52 \\
\bottomrule
\label{tab:ranking}
\end{tabular}
\end{adjustbox}
\vspace{-7mm}
\end{table}

\vspace{-3mm}
\subsection{Speech Quality Assessment}
We evaluate NOMAD for speech quality assessment using Pearson's correlation coefficient (PC) and SC of the NOMAD score against MOS. We consider 4 different speech MOS databases that cover a broad range of degradations. The ITU-T Supplement 23 to the P series of the ITU-T Recommendations Experiment 1 (P23 EXP1) and Experiment 3 (P23 EXP3)~\cite{ITUT1998} are used to evaluate various codecs and an 8 kbps codec under different channel degradations respectively. The TCD-VoIP database is used to test typical degradations occurring in VoIP communications~\cite{harte2015tcd}. The Genspeech database includes parametric and generative codecs presenting differences such as slight pitch shift and microalignments which are imperceptible but penalized by full-reference metrics (ViSQOL, PESQ)~\cite{jassim2021warp}. 
The results aggregated per condition (Table \ref{tab:MOS_pred}) show that NOMAD outperforms both NORESQA and the wav2vec 2.0 features and exhibits competitive results with full-reference metrics. Our method shows high invariance to clean speech demonstrated by the very close correlation scores between the non-matching reference NOMAD version and the full-reference mode (NOMAD FR) where we only used the clean counterpart as a reference.

\vspace{-2mm}
\begin{table}[!t]
\caption[C]{PC and SC of non-matching reference (NMR) and full-reference (FR) models.}
\centering
\Huge
\ra{1.1}
\begin{adjustbox}{max width=0.45\textwidth}
\fontsize{24pt}{24pt}\selectfont
\begin{tabular}{@{}lllcccccccccccccccrrrrrrrrrrrrrrrrrr@{}}\toprule

& \multicolumn{2}{c} {\phantom} & \multicolumn{4}{c} {\textbf{P23 EXP 1}} & \multicolumn{4}{c}{\textbf{P23 EXP 3}} & \multicolumn{4}{c}{\textbf{TCD-VoIP}} & \multicolumn{4}{c}{\textbf{GENSPEECH}}\\
\cmidrule{4-6} \cmidrule{8-10} \cmidrule{12-14} \cmidrule{16-18}

\textbf{Type} && \textbf{Model} &  PC && SC && PC && SC && PC && SC && PC && SC &                                \\ \midrule
NMR && NOMAD       & \textbf{-0.85} && \textbf{-0.88}  && \textbf{-0.85} && \textbf{-0.75}  && \textbf{-0.64} && \textbf{-0.64} && \textbf{-0.94} && \textbf{-0.90}    \\
 && w2v     & -0.26 && -0.27  && -0.38 && -0.36  && -0.39 && -0.54 && -0.67 && -0.90    \\
  && NORESQA & -0.24 && -0.20  && -0.46 && -0.23  && -0.11 && -0.14 && -0.69 && -0.69    \\

\hline
FR &&NOMAD FR    & -0.86 && -0.87  && -0.86 && -0.73 && -0.63 && -0.65 && \textbf{-0.96} && \textbf{-0.90}       \\
&&CDPAM       & -0.48 && -0.35  && -0.39 && -0.37 && -0.76 && -0.79 && -0.93 && \textbf{-0.90}       \\
&&ViSQOL      & 0.87  && 0.89   && 0.78  && 0.67 && 0.74 && 0.76 && 0.64 && 0.74       \\
&&WARP-Q      & -0.88 && -0.92  && \textbf{-0.87} && -0.79 && -0.90 && \textbf{-0.92} && -0.89 && \textbf{-0.90} \\
&&PESQ        & \textbf{0.91}  && \textbf {0.96}   && \textbf{0.87} && \textbf{0.87} && \textbf{0.91} && 0.91 && 0.49 && 0.52       \\
\bottomrule
\label{tab:MOS_pred}
\end{tabular}
\end{adjustbox}
\vspace{-4mm}
\end{table}

\subsection{Speech Enhancement}
We evaluate NOMAD loss for the speech enhancement task using the model DEMUCS~\cite{defossez2020real} following a similar approach of~\cite{manocha2021cdpam}. We train three models using the Valentini speech dataset (28 speakers)~\cite{valentini}; (1) The original \textit{DEMUCS} trained from scratch using L1 loss between waveforms and multi-resolution STFT~\cite{defossez2020real}; (2) \textit{MT NOMAD} combines the losses of DEMUCS with the NOMAD loss in a multitask fashion; (3) \textit{FT NOMAD} is based on finetuning the pretrained DEMUCS model in (1) using the NOMAD loss only. The NOMAD loss is computed as the sum of the L1 distance between clean speech and the estimated speech of each transformer layer and the embedding layer for every time frame. The frame-wise approach is preferred for this task to encourage a local reconstruction that might be lost in the final embedding layer.
Every model is trained for 110 epochs with batch size set to 8. For testing, the best model is taken as the one with the lowest validation loss. The validation partition is created by leaving out 2 speakers from the Valentini training set as mentioned in the DEMUCS repo~\cite{defossez2020real}. Results are evaluated on the Valentini test set with PESQ and a listening test. PESQ is computed on the entire test set which includes 824 noisy speech samples at four SNR values 2.5, 7.5, 12.5, 17.5 dB, 1 male and 1 female speaker, and 5 noise types.
A MUSHRA test is conducted with 12 samples, distributed in 4 recordings for 3 SNR values. For each SNR we take 2 male speaker and 2 female speaker samples and 4 noise types. In each MUSHRA session, we use 5 stimuli: noisy sample (anchor), clean (hidden reference), and three enhanced versions from DEMUCS, FT NOMAD, and MT NOMAD respectively. Listeners could also play the clean reference to compare. 
We recruited 16 people for the listening test using the online platform Go listen~\cite{barry2021go}. We asked raters to indicate their knowledge of audio as follows; 80\% as professionals working in the area of audio, 20\% as audio enthusiasts, and 0\% rarely paying attention to audio quality. Post-screening was done as indicated in the MUSHRA guidelines~\cite{MUSHRA}. We removed 3 participants who judged the hidden reference under 90 for more than 15\% of samples. Further, we removed another participant (audio enthusiast) who scored 0 on all enhanced models.   
In Table \ref{tab:se_results}, for each SNR value we report the average PESQ, and the median and interquartile range for the MUSHRA test as recommended in \cite{MUSHRA}. Results indicate that both approaches improve over the baseline for both metrics. An inconsistency can be noted between subjective and objective scores i.e., MT NOMAD exhibits the highest MUSHRA scores while FT NOMAD shows the highest PESQ scores. The MOS predictions from PESQ for the samples used in the subjective study had a high correlation (SC=0.89) with the subjective scores.
This speech enhancement study further demonstrates that NOMAD embeddings encode perceptual similarity and that they can also be applied to generative tasks. A potential future application to further showcase the versatility of NOMAD embeddings is in non-parallel speech enhancement, where any clean signal can serve as the ground truth.

\begin{table}[!t]
\caption[C]{Speech enhancement performance evaluation.}
\centering
\ra{1.0}
\begin{adjustbox}{max width=0.49\textwidth}
\fontsize{14pt}{14pt}\selectfont
\begin{tabular}{@{}lccccccc|ccccccccccccccccccccc@{}}\toprule

& \multicolumn{7}{c} {\textbf{PESQ}$\uparrow$}  \vline & \multicolumn{6}{c}{\textbf{MUSHRA}$\uparrow$} \\

&  2.5 && 7.5 && 12.5 && 17.5 && 2.5 && 7.5 && 12.5  \\ \midrule
Noisy         & 1.42 && 1.76 && 2.10 && 2.60 && 20 \textit{(10,52)} && 30 \textit{(19,65)} && 46 \textit{(20,82)} \\
Demucs (\textit{Baseline})       & 2.40 && 2.83 && 3.06 && 3.31 && 58 \textit{(39,78)} && 78 \textit{(51,90)} && 84 \textit{(58,90)}   \\
FT Nomad (\textit{Ours})      & \textbf{2.43} && \textbf{2.88} && \textbf{3.14} && \textbf{3.42} && 70 \textit{(50,82)} && 80 \textit{(50,90)} && 88 \textit{(57,91)} \\
MT Nomad (\textit{Ours})     & 2.42 && 2.84 && 3.10 && 3.36 && \textbf{72} \textit{(58,84)} && \textbf{90} \textit{(63,94)} && \textbf{90} \textit{(71,95)} \\
 \bottomrule
\label{tab:se_results}
\end{tabular}
\end{adjustbox}
\vspace{-5mm}
\end{table}

\vspace{-7mm}
\section{Conclusions}
We proposed NOMAD, a non-matching reference perceptual similarity metric that can be used for perceptual audio tasks. Future work will further analyse the role of wav2vec 2.0 in NOMAD. Its use is supported by its capacity to disentangle variational factors in speech and its superior performance compared to a model we trained from scratch. NOMAD outperforms other models in the task of ranking degradations and audio quality prediction with non-matching clean references. We observe that the fixed dimension of NOMAD embeddings helps in solving issues of microalignment of generative neural codecs, which is a known problem of full-reference metrics (ViSQOL, PESQ). Objective and subjective experiments show that NOMAD can be used as a perceptual loss for speech enhancement to further improve speech quality. Beyond the evaluated tasks, we believe that the proposed model could be used for many other generative tasks such as text-to-speech, as a feature extractor for no-reference quality metrics and to measure quality relative to any reference chosen. 

\bibliographystyle{IEEEbib}
\bibliography{strings,refs}
\end{document}